\documentclass[aps,prd,groupedaddress,amsmath,amssymb,amsfonts,nofootinbib,preprint]{revtex4-1}

\usepackage{amssymb,amsfonts}
\usepackage{mathrsfs}
\usepackage{tikz}
\usetikzlibrary{external}

\usepackage{graphicx}
\usepackage[utf8x]{inputenc}
\usepackage{float}
\usepackage{amsmath}
\DeclareMathAlphabet{\mathpzc}{OT1}{pzc}{m}{it}
\usetikzlibrary{patterns}

\begin{document}

\bibliographystyle{plain}

\title{Memory effect for particle scattering in odd spacetime dimensions}

\author{Gautam Satishchandran}
\email{gautamsatish@uchicago.edu}
\author{Robert M. Wald}
\email{rmwa@uchicago.edu}
\affiliation{Enrico Fermi Institute and Department of Physics \\
  The University of Chicago \\
  5640 South Ellis Avenue, Chicago, Illinois 60637, USA.}

\begin{abstract}
We investigate the gravitational memory effect for linearized perturbations off of Minkowski space in odd spacetime dimensions $d$ by examining the effects of gravitational radiation from classical point particle scattering. We also investigate analogous memory effects for electromagnetic and scalar radiation. We find that there is no gravitational memory effect in all odd dimensions. For scalar and electromagnetic fields, there is no memory effect for $d\geq 7$; for $d=3$ there is an infinite momentum memory effect, whereas for $d=5$ there is no momentum memory effect but the displacement of a test particle will grow unboundedly with time. Our results are further elucidated by analyzing the memory effect for any slowly moving source of compact spatial support in odd dimensions. 

\end{abstract}

\maketitle
\section{Introduction}
Gravitational waves induce relative displacements in a system of test particles. Within the context of linearized gravity in four spacetime dimensions, it has been known for more than 40 years \cite{Zeldovich} that a pulse of gravitational radiation can cause a permanent change in the relative displacement of these particles. More generally, in asymptotically flat four-dimensional spacetimes, such a permanent displacement in the relative position of test particles near null infinity will occur quite generally in the presence of gravitational radiation at the same order (namely, $1/r$) as other effects of gravitational radiation. This effect is known as the ``memory effect'' and is currently an active area of research \cite{Strominger,{Pasterski,Ashtekar,Flanagan}}.

Recently, it was shown \cite{Hollands,Tolish1} that the memory effect vanishes in higher, even-dimensional spacetimes; i.e., for $d$-dimensional spacetimes with $d>4$ and even, there is no permanent displacement of test particles near null infinity occurring at the same leading order as the effects of gravitational radiation (namely, at order $1/r^{d/2 -1}$). The analysis of \cite{Hollands} obtained this result by considering the general behavior of the metric near null infinity with the assumption that the spacetime becomes stationary near spatial infinity and at sufficiently late times. The analysis of \cite{Tolish1} obtained this result by considering classical point particle scattering in linearized gravity in $d$-dimensional Minkowski spacetime. 

The analysis of \cite{Hollands} was restricted to even-dimensional spacetimes because conformal null infinity does not exist for radiating solutions in odd dimensions \cite{Ishibashi,Hollands2}. Thus, the methods of \cite{Hollands} cannot be applied. The analysis of \cite{Tolish1} was restricted to even dimensions because the retarded Green's function for the wave equation in $d$-dimensional Minkowski spacetime differs markedly when $d$ is odd from the even-dimensional case. In even dimensions, the retarded Green's function satisfies Huygens' principle; i.e., the support of the Green's function lies entirely on the light cone. By contrast, the retarded Green's function to the wave equation in odd dimensions has additional support in the interior of the light cone (see, e.g., \cite{Friedlander}). Thus, in odd dimensions, even for a scattering event that is highly localized in time, the radiative effects will persist over all time and the memory effects cannot be deduced as straightforwardly as in the even-dimensional case.

The purpose of this paper is to extend the analysis of \cite{Tolish1} to the odd-dimensional case for spacetime dimension\footnote{For $d=3$, there is no gravitational radiation and thus, obviously, no gravitational memory effect, but scalar and electromagnetic radiation exist for $d=3$.} $d \geq 3$. We calculate the retarded field resulting from classical particle scattering in Minkowski spacetime as in \cite{Tolish1} for the scalar, electromagnetic, and linearized gravity cases. Specifically, we consider a source consisting of incoming particles that move with uniform velocity (i.e., on geodesics of the Minkowski metric) and meet at a single event $O$. At $O$, the particles may interact, destroy each other and/or create new particles, subject only to conservation of charge at $O$ in the electromagnetic case and conservation of $4$-momentum at $O$ in the gravitational case. The outgoing particles then move inertially to future infinity. 

Our main result for the linearized gravity case is as follows: We calculate the retarded metric perturbation for such a particle scattering event, keeping the leading order term in an expansion in $1/r$. For $d \geq 5$, the corresponding leading order behavior of the Riemann tensor\footnote{The Riemann tensor vanishes outside of the source for $d=3$.} is found to be [see \eqref{Riem} below]
\begin{equation}
R_{abcd} \propto \frac{1}{r^{d/2 - 1}} \bigg(\frac{\partial}{\partial U}\bigg)^{\frac{d-1}{2}}\bigg(\frac{\Theta(U-U_{O})}{\sqrt{U-U_{O}}}\bigg)
\label{riem0}
\end{equation}
where $U = t - r$ is the retarded time and $U_{O}$ is the retarded time of the interaction event $O$. 
This Riemann tensor \eqref{riem0} produces a singular effect on the relative motion of test particles near time $U_{O}$, which can be understood as an artifact of our use of idealized point particle sources that interact at a single event $O$. If we suitably ``smooth out'' the source so that the interaction takes place over a finite time interval, we find that the relative displacement of the particles oscillate over the interaction time scale and then relax back to the original displacement. Thus, there is no memory effect.

In Sec.II we use the retarded Green's function for the wave equation in odd-dimensional Minkowski spacetime to obtain the retarded solution for scalar point charge scattering. In Sec.III, we compute the scalar and electromagnetic analogs of memory. In Sec.IV we compute the retarded gravitational field and analyze the associated memory effect, as described above. In Sec.V we analyze memory in the slow motion limit for an arbitrary source.

We work in geometrized units $(G=c=1)$. We will use the notation and sign conventions of \cite{Wald}. In particular, our metric signature is ``mostly positive'' and our sign convention for curvature is such that the scalar curvature of a round sphere is positive. Latin indices from the early alphabet ($a,b, \dots$) denote abstract spacetime indices. Greek indices ($\mu, \nu, \dots$) denote spacetime components of tensors, whereas latin indices from the middle alphabet ($i,j, \dots$) denote spatial components. If $f$ is a distribution, we denote its $k$th distributional derivative as $f^{(k)}$. The $k$-dimensional delta function will be denoted as $\delta_{k}$. 

\section{Retarded Solution to the Scalar Wave Equation For an Arbitrary Particle Interaction}

Consider the inhomogeneous wave equation for a field $\phi$ sourced by a charge density $S$ on $d$-dimensional Minkowski spacetime $(\mathbb{R}^{d},\eta_{ab}):$ 
\begin{equation}
\eta^{ab}\partial_{a}\partial_{b}\phi = -4\pi S.
\label{waveeq}
\end{equation}
When $d \geq 3$ is odd, the retarded Green's function for this equation takes the form \cite{Friedlander}
\begin{equation}
G(x,x') = \lim_{\epsilon\downarrow 0} \frac{\Theta(t-t')}{2\pi^{\frac{d-1}{2}}}\zeta^{\big(\frac{d-3}{2}\big)}(\sigma-\epsilon).
\label{Greensfunctionodd}
\end{equation}
Here $\zeta$ is the distribution on $\mathbb{R}$ given by
\begin{equation}
\zeta(x) \equiv \frac{\Theta(x)}{\sqrt{x}}
\end{equation}
and $\zeta^{\big(\frac{d-3}{2}\big)}$ denotes the $(d-3)/2$ distributional derivative of $\zeta$ as a distribution on $\mathbb{R}$. The quantity $\sigma$ is minus the squared geodesic distance between $x$ and $x'$, i.e.,
\begin{equation}
\sigma = (t-t')^{2}-|\mathbf{x}-\mathbf{x}'|^2.
\end{equation}
The $\epsilon$ has been inserted in \eqref{Greensfunctionodd} because the composition $\zeta^{\big(\frac{d-3}{2}\big)} \circ \sigma$ is not manifestly well defined as a distribution on $\mathbb{R}^d \times \mathbb{R}^d$ on account of the fact that when $x = x'$ we have both $\sigma = 0$ (so $\zeta$ and its derivatives are singular) and $\nabla_{a} \sigma = 0$ (so the level sets of $\sigma$ are not well behaved). However, when $\epsilon > 0$, $\zeta^{\big(\frac{d-3}{2}\big)} \circ \sigma_\epsilon$ with $\sigma_\epsilon \equiv \sigma - \epsilon$ is manifestly well defined as a distribution on spacetime, as can be seen from the fact that $\sigma_{\epsilon}$ could be used as a coordinate covering a neighborhood in $\mathbb{R}^d \times \mathbb{R}^d$ of the region where $\sigma = \epsilon$. One can verify directly that the limit as $\epsilon \downarrow 0$ appearing
on the right side of \eqref{Greensfunctionodd} exists and defines a distribution on $\mathbb{R}^d \times \mathbb{R}^d$ that satisfies 
\begin{equation}
\eta^{ab}\partial_{a}\partial_{b}G(x,x')=-\delta(x,x').
\end{equation}
Since $G(x,x')$ vanishes whenever $x$ is not in the causal future of $x'$, it therefore is the retarded Green's function for the wave equation \eqref{waveeq}.

We can rewrite \eqref{Greensfunctionodd} in a useful form as follows. If $D$ is any distribution on $\mathbb{R}$ and $F: M \rightarrow \mathbb{R}$ is any smooth function on a manifold $M$ with $\nabla_a F \neq 0$, then $D \circ F$ is a distribution on $M$. Its (distributional) gradient on $M$ is given by
\begin{equation}
\nabla_a (D \circ F) = (\nabla_a F) D^{(1)} \circ F
\end{equation}
where $D^{(1)}$ is the derivative of $D$ as a distribution on $\mathbb{R}$. If $\xi^a$ is a vector field on $M$ with $\xi^a \nabla_a F \neq 0$ in the support of $D$, then we have
\begin{equation}
D^{(1)} \circ F = \frac{1}{\xi^a \nabla_a F }\xi^a \nabla_a (D \circ F).
\end{equation}
Applying this formula for $D=\zeta$ and $F=\sigma_\epsilon = \sigma - \epsilon$, and choosing 
$\xi^{a'} =(\partial/\partial t')^{a'}$, we obtain
\begin{equation}
G(x,x') = \lim_{\epsilon\downarrow 0} \frac{1}{(2\pi)^{\frac{d-1}{2}}}\bigg(-\frac{1}{t-t'}\frac{\partial}{\partial t'}\bigg)^{\frac{d-3}{2}}\bigg(\frac{\Theta(t-t'-|\mathbf{x}-\mathbf{x}'| - \epsilon)}{\sqrt{(t-t')^{2}-|\mathbf{x}-\mathbf{x}'|^{2} - \epsilon}}\bigg)
\label{Greensfunctiontimederiv}
\end{equation}
where the derivative is now taken in the spacetime distributional sense. 

\begin{figure}
\centering
\includegraphics[scale=1.1]{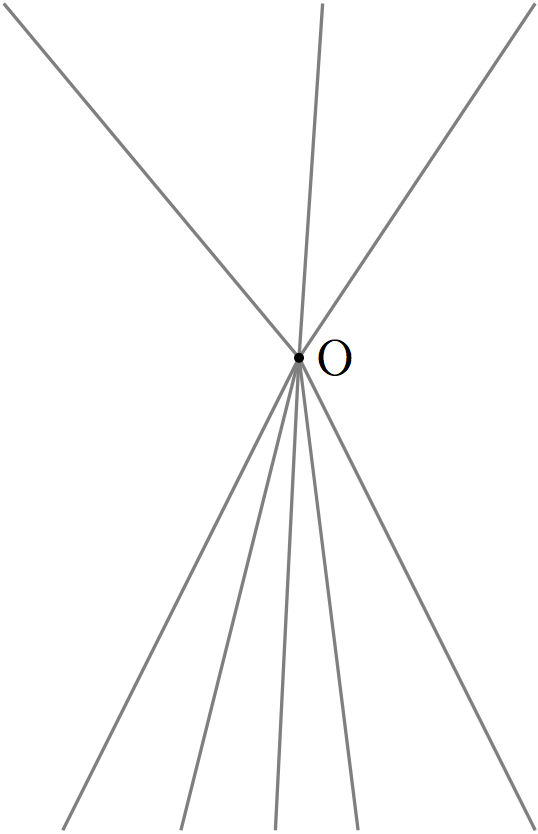}
\caption{A spacetime diagram representative of the type of particle scattering we consider. Here we have five particles scattering at event $O$ and three particles emerge. The worldlines of the incoming and outgoing particles are timelike.}
\label{scatter}
\end{figure}

The retarded solution to (\ref{waveeq}) is given by 
\begin{equation}
\phi_S(x) = 4\pi\int{G(x,x')S(x')d^{d}x'}.
\label{phiintegralexp}
\end{equation}
As already mentioned in the Introduction, we will take $S$ to correspond to a system of incoming point particles with scalar charge that move on timelike inertial worldlines that intersect at event $O$, as illustrated in Fig.\ref{scatter}. At $O$, the particles may interact and/or be created or destroyed. The outgoing particles also move on timelike geodesics. Since there is no conservation law for scalar charge, there is no restriction on the charges of the incoming or outgoing particles.

Without loss of generality we choose a global inertial coordinate system $(t,\textbf{x})$ so that $O$ lies at the origin. The source $S$ is then given by 
\begin{equation}
S(x)=\sum_{(j)in}q^{\rm in}_{(j)}\frac{d\tau_{(j)}}{dt}\delta_{d-1}(\mathbf{x}-\mathbf{y}_{(j)}(t))\Theta(-t) + \sum_{(i)out}q^{\rm out}_{(i)}\frac{d\tau_{(i)}}{dt}\delta_{d-1}(\mathbf{x}-\mathbf{y}_{(i)}(t))\Theta(t)\label{gensource}
\end{equation}
where $q^{\rm in}_{j}$ $(q^{\rm out}_{i})$ are the in (out) charges as measured in their rest frame, $\textbf{y}_{(i)}(t)$ is the spatial trajectory of the $i$th particle as a function of $t$ and $\tau_{(i)}$ is the proper time along the worldline of the $i$th particle.

We wish to evaluate \eqref{phiintegralexp} for the source \eqref{gensource}. To begin, consider the simple case where there are no incoming particles and a single particle is created at rest at $O$. The source is then simply
\begin{equation}
S_0=q\delta_{d-1}(\mathbf{x})\Theta(t).
\label{createdscalar}
\end{equation}
Substituting in \eqref{phiintegralexp} and using the form \eqref{Greensfunctiontimederiv} of the retarded Green's function, we obtain the retarded solution
\begin{equation}
\phi_0(x)= \frac{4\pi q}{(2\pi)^{\frac{d-1}{2}}}\int_0^\infty{dt'\bigg(-\frac{1}{t-t'}\frac{\partial}{\partial t'}\bigg)^{\frac{d-3}{2}}\bigg(\frac{\Theta(t-t'-r)}{\sqrt{(t-t')^{2}-r^{2}}}\bigg)}
\label{phi0}
\end{equation}
where we have written $r=|\mathbf{x}|$. The ``$\epsilon$-regularization'' of the Green's function (see \eqref{Greensfunctiontimederiv}) is not needed here or in the equations below since we are interested in evaluating the solution only for $r > 0$, i.e., outside of the source. Clearly, $\phi_0 = 0$ whenever $t < r$. 

For $d=3$, we find that when $t \geq r$, we have
\begin{eqnarray}
\phi_0(x) &=& 2q \int_0^{t - r} dt' \frac{1}{\sqrt{(t-t')^{2}-r^{2}}}  \nonumber \\
&=& 2q \cosh^{-1} (t/r).
\end{eqnarray}
Interestingly, at fixed $\mathbf{x}$, $\phi_0$ diverges logarithmically in $t$ as $t \rightarrow \infty$ and does not approach the static solution $-2q \ln r$. On the other hand, for $t > r$, we have
\begin{equation}
\nabla_a \phi_0(x)= - \frac{2q}{\sqrt{t^2-r^2}}\left(-\nabla_a t  +  \frac{t}{r} \nabla_a r \right)
\label{statlim}
\end{equation}
which does approach the gradient of the static solution $\nabla_a(-2q \ln r) = -2q/r \nabla_a r$ as $t \to \infty$ at fixed $\mathbf{x}$.

For $d \geq 5$, we can obtain the leading order behavior in $1/r$ of $\phi_0$ as follows. Consider, first, the case $d=5$. Writing $U = t - r$, we obtain from \eqref{phi0} 
\begin{equation}
\phi_0(x)= -\frac{q}{\pi}\int_0^\infty{dt' \frac{1}{U+r-t'}\frac{\partial}{\partial t'} \bigg(\frac{\Theta(U-t')}{\sqrt{(U+r-t')^{2}-r^{2}}}\bigg)}.
\end{equation}
We are interested in the behavior of $\phi_0$ as $r \to \infty$ at fixed $U > 0$. Integrating by parts, we obtain
\begin{eqnarray}
\phi_0(x)= && \frac{q}{\pi}\int_0^\infty{dt' \frac{\partial}{\partial t'}\bigg(\frac{1}{U+r-t'}\bigg) \frac{\Theta(U-t')}{\sqrt{(U+r-t')^{2}-r^{2}}}} \nonumber \\
&-& \frac{q}{\pi}\frac{1}{U+r-t'}\frac{\Theta(U-t')}{\sqrt{(U+r-t')^2 - r^2}} \bigg\vert^\infty_0.
\end{eqnarray}
It is easily seen that the first term decays as $r^{-5/2}$ as $r \to \infty$ at fixed $U$. In the second term, the boundary term from infinity clearly vanishes. However, the boundary term from $0$ is nonvanishing, and decays only as $r^{-3/2}$. Thus, this term dominates the behavior of $\phi_0$ as $r \to \infty$ at fixed $U$, and we find
\begin{equation}
\phi_0 = \frac{q}{\sqrt{2} \pi r^{3/2}} \frac{\Theta(U)}{\sqrt{U}} +O(r^{-5/2}).
\end{equation}
Repeating this calculation for general odd $d$ with $d \geq 5$, we obtain
\begin{equation}
\phi_{0} = \frac{2\sqrt{\pi}q}{(2\pi r)^{\frac{d}{2}-1}}\bigg(\frac{\partial}{\partial U}\bigg)^{\frac{d-5}{2}}\bigg(\frac{\Theta(U)}{\sqrt{U}}\bigg) + O(1/r^{d/2})
\label{phiout0}
\end{equation}
where the derivative of the spacetime distribution $\Theta(U)/\sqrt{U}$ is meant in the distributional sense.

The field of a particle created at $O$ with a general inertial worldline $(t,\mathbf{y}(t))$ can be obtained by boosting (\ref{phiout0}). We obtain
\begin{equation}
\phi_{0,v}(x)=\frac{2\sqrt{\pi}q}{(2\pi r)^{\frac{d}{2}-1}}\frac{d\tau}{dt}\bigg(\frac{1}{1-\hat{\mathbf{r}}\cdot\mathbf{v}}\bigg)\bigg(\frac{\partial}{\partial U}\bigg)^{\frac{d-5}{2}}\bigg(\frac{\Theta(U)}{\sqrt{U}}\bigg) + O(1/r^{d/2})
\label{phioutv}
\end{equation}
where $\tau$ is the proper time along the worldline, $\mathbf{v}$ is the coordinate velocity $\mathbf{v}=d\mathbf{y}/dt$, and $\hat{\mathbf{r}}=\mathbf{x}/r$. 

Consider, now, a particle that is at rest at the origin until $t=0$, at which time it is destroyed, i.e.,
\begin{equation}
\tilde{S}_0=q\delta_{d-1}(\mathbf{x})\Theta(-t).\label{destroyedscalar}
\end{equation}
For $d=3$, the retarded solution is given by
\begin{equation}
\tilde{\phi}_0 = 2q \int_{-\infty}^\alpha dt' \frac{1}{\sqrt{(t-t')^{2}-r^{2}}}
\end{equation}
where $\alpha = {\rm min} (0, t - r)$. However, this integral clearly does not converge, so the retarded solution for a particle at rest in the infinite past does not exist. However, if we worked with $\nabla_a 
\tilde{\phi}_0$ rather than $\tilde{\phi}_0$, then the corresponding retarded integral would converge and would yield the gradient of the static solution for $t < r$.

On the other hand, the retarded integral for $\tilde{\phi}_0$ does converge when $d \geq 5$ and, for $t - r < 0$, it agrees with the solution, $ \phi_{\rm static}$, for a static particle [i.e., the solution to the Poisson equation in $(d-1)$ dimensions with a $\delta$-function source]. It follows that for $d \geq 5$, we have
\begin{equation}
\tilde{\phi}_0 = \phi_{\rm static} - \phi_0
\end{equation}
where $\phi_0$ is given by \eqref{phiout0}. Since $\phi_{\rm static}$ in $d-1$ spatial dimensions falls off as $1/r^{(d-3)}$, we see that the leading order behavior in $1/r$ of $\tilde{\phi}_0$ is just minus that of $\phi_0$. This result clearly also holds for boosted particles.

Putting the above results together, we see that for $d \geq 5$, the leading order behavior of the retarded solution, $\phi_S$, with source 
\eqref{gensource} is given by 
\begin{equation}
\phi_{S}(x)=\frac{2\sqrt{\pi}(\alpha(\hat{\mathbf{r}})-\beta(\hat{\mathbf{r}}))}{(2\pi r)^{\frac{d}{2}-1}}\bigg(\frac{\partial}{\partial U}\bigg)^{\frac{d-5}{2}}\bigg(\frac{\Theta(U)}{\sqrt{U}}\bigg)
\label{phiSv}
\end{equation}
where
\begin{equation}
\begin{aligned}
\alpha(\hat{\mathbf{r}})=\sum_{(i)out}q^{\rm out}_{(i)}\frac{d\tau_{(i)}}{dt}\bigg(\frac{1}{1-\hat{\mathbf{r}}\cdot\mathbf{v}_{(i)}}\bigg)\\
\beta(\hat{\mathbf{r}})=\sum_{(j)in}q^{\rm in}_{(j)}\frac{d\tau_{(j)}}{dt}\bigg(\frac{1}{1-\hat{\mathbf{r}}\cdot\mathbf{v}_{(j)}}\bigg).\\
\end{aligned}
\label{alphabeta}
\end{equation}
When $d=3$, the retarded solution with source \eqref{gensource} does not, in general, exist, but an expression analogous to \eqref{phiSv} could be given for $\nabla_a \phi_{S}$.

\section{The Scalar and Electromagnetic Memory Effect}\label{scandelmem}

In this section, we examine the effects of the retarded scalar field solution, $\phi_S$, with source \eqref{gensource} on a distant scalar charged test particle. We then perform a similar analysis in the electromagnetic case. Our main interest is to investigate whether, to leading nontrivial order in $1/r$, there is any permanent change in the position or momentum of a test particle. The corresponding analysis for the gravitational case will be performed in the next section. 

\subsection{Scalar memory}

The force on a test particle of scalar charge $\mathcal{Q}$ in an external scalar field $\phi$ is
\begin{equation}
f^{a}= \mathcal{Q} \nabla^a \phi.
\end{equation}
Thus, for $d \geq 5$, it follows directly from \eqref{phiSv} that, to leading order, the force on a test particle of scalar charge $\mathcal{Q}$ at large distances from the source \eqref{gensource} is
\begin{equation}
f^{a}=-\frac{\mathcal{Q}2\sqrt{\pi}(\alpha-\beta)}{(2\pi r)^{\frac{d}{2}-1}}\bigg (\frac{\partial}{\partial U}\bigg)^{\frac{d-3}{2}} \bigg(\frac{\Theta(U)}{\sqrt{U}}\bigg)K^{a}
\label{scalarforce}
\end{equation}
where $K^{a}=-\partial^{a}U$ and $\alpha$ and $\beta$ are given by (\ref{alphabeta}). Using \eqref{statlim} and its generalization to boosted particles, one may verify that this formula also holds for $d=3$. 

For $d=3$, the force decays in time only as $1/\sqrt{U}$, so for any $U_0$, the integrated effect of the force on the momentum for all $U > U_0$ is never negligible. In this sense, there is an infinite momentum memory effect for $d=3$. We will restrict consideration to $d \geq 5$ for the remainder of this subsection.

If the test particle motion is nonrelativistic (so that proper time, $\tau$, along its worldline agrees with retarded time $U$), then the change in momentum is given by 
\begin{equation}
(\Delta P)^{a}(U)=\int_{-\infty} ^{U}{dU'f^{a}(U')}=-\frac{\mathcal{Q}2\sqrt{\pi}(\alpha-\beta)}{(2\pi r)^{\frac{d}{2}-1}} \bigg (\frac{\partial}{\partial U}\bigg)^{\frac{d-5}{2}} \bigg(\frac{\Theta(U)}{\sqrt{U}}\bigg)K^{a}.
\label{scalarmom}
\end{equation}
As $U \downarrow 0$, $(\Delta P)^{a}$ becomes arbitrarily large, which is not consistent with our assumption of nonrelativistic motion. However, this blowup of $(\Delta P)^{a}$ for $U$ near $0$ is an artifact of our idealization of the particle interactions as taking place at a sharply defined event $O$.
For a smoothed out source, the interaction would occur over some finite time, and for sufficiently large $r$, the change in $4$-momentum would be small at all times. For example, we could smear the source slightly in time by considering the new source
\begin{equation}
S'(t,{\mathbf x}) = \int dt' \chi (t-t') S(t',{\mathbf x}) 
\end{equation}
where $\chi(\xi)$ is non-negative function that is sharply peaked around $\xi = 0$ and satisfies $\int d\xi \chi(\xi) = 1$. By linearity and time translation symmetry, the retarded solution with source $S'$
is obtained by convolving $\phi_S$ [see \eqref{phiSv}] with $\chi$. To leading order in $1/r$, the momentum change of a test particle then becomes
\begin{eqnarray}
(\Delta P)^{a}_{\rm smooth}(U) &=&-\frac{\mathcal{Q}2\sqrt{\pi}(\alpha-\beta)K^{a}}{(2\pi r)^{\frac{d}{2}-1}}\int_{-\infty}^{\infty}{d\xi\bigg (-\frac{\partial}{\partial \xi}\bigg)^{\frac{d-5}{2}} \bigg(\frac{\Theta(U-\xi)}{\sqrt{U-\xi}}\bigg)\chi(\xi)} \nonumber \\
&=& -\frac{\mathcal{Q}2\sqrt{\pi}(\alpha-\beta)K^{a}}{(2\pi r)^{\frac{d}{2}-1}}\int_{-\infty}^{U}{d\xi
\frac{1}{\sqrt{U-\xi}} 
\bigg (\frac{\partial}{\partial \xi}\bigg)^{\frac{d-5}{2}} \chi(\xi)}
\label{mommem}
\end{eqnarray}
which is finite for all $U$. 
It is easily seen from \eqref{mommem} that for all $d\geq 5$, we have $(\Delta P)^{a}_{\rm smooth}(U) \to 0$ as $U \to \infty$, so there is no ``momentum memory effect.'' 

\begin{figure}
\centering
\includegraphics[scale=1.5]{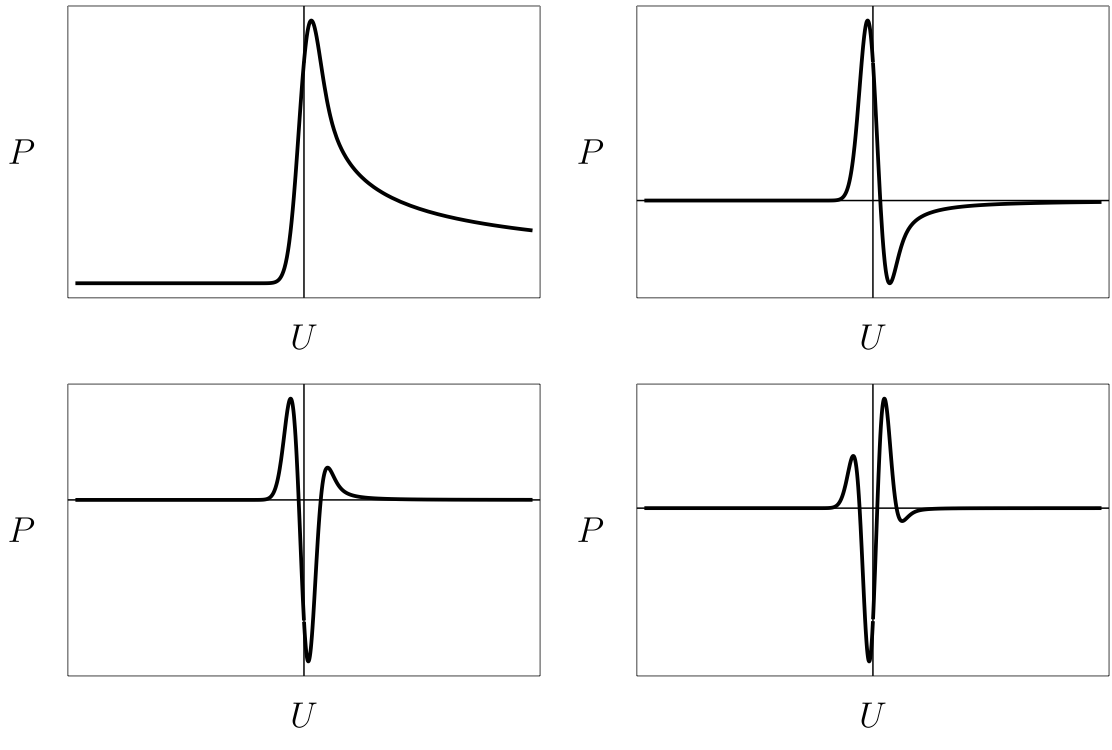}
\caption{The change in momentum of a scalar charge, initially at rest, as a function of retarded time in 5 (top left), 7 (top right), 9 (lower left) and 11 (lower right) dimensions after smearing out the particle interaction event with a Gaussian normalized in time. The test charge's momentum generically oscillates over the time scale of the interaction $(\tau)$ and then relaxes to its initial value. In this figure, $\tau$ is taken to be much smaller than the range of $U$ plotted.}\label{momdetector}
\end{figure}

Figure \ref{momdetector} illustrates the behavior of $(\Delta P)^{a}_{\rm smooth}$ in various dimensions for the case where $\chi$ is chosen to be a normalized Gaussian in time, i.e., $\chi(\xi) = \sqrt{\pi}\tau\exp(-\xi^2/\tau^2)$.
We find that the change in momentum will oscillate about the particles' initial momentum $(d-5)/2$ times over a time scale $\tau$, with an amplitude of the oscillation that scales as $\tau^{-\big(\frac{d}{2}-2\big)}$. $(\Delta P)^{a}_{\rm smooth}$ then relaxes back to zero as $U^{-\big(\frac{d}{2}-2\big)}$. 

The displacement of the test particle is obtained by integrating its momentum in time (again, assuming nonrelativistic motion of the test particle). For $d=5$, the displacement will grow as $\sqrt{U}$ for large $U$. In this sense, for $d=5$ there is an infinite displacement memory effect. For $d > 5$, the integral of the momentum will converge to zero at large $U$, so there is no displacement memory effect.

\subsection{Electromagnetic memory}

Maxwell's equations for the vector potential $A^{a}$, in the Lorenz gauge $(\nabla_{a}A^{a}=0)$, take the form of a wave equation on $(\mathbb{R}^{d},\eta_{ab})$:
\begin{equation}
\eta^{ab}\partial_{a}\partial_{b}{A^{c}}=-4\pi J^{c}. \label{elwaveeq}
\end{equation}
It follows immediately that the Green's function for \eqref{elwaveeq} is given by
\begin{equation}
G^{a}{}_{a'}(x,x') = \eta^{a}{}_{a'} G(x,x')
\end{equation}
where $G(x,x')$ is the Green's function to \eqref{waveeq} and $\eta^{a}{}_{a'}$ is the parallel propagator, which is trivial in global inertial coordinates.

For a scattering event of the type pictured in Fig.\ref{scatter}, the current density of 
the outgoing particles in the electromagnetic case takes the form
\begin{equation}
J^{a}_{(out)}=\sum_{i,out}q^{\rm out}_{(i)}\frac{d\tau_{(i)}}{dt}u_{(i)}^{a}\delta_{d-1}(\mathbf{x}-\mathbf{y}_{(i)}(t))\Theta(t)
\end{equation}
and the current density of the incoming particles takes the form
\begin{equation}
J^{a}_{(in)}=\sum_{j,in}q^{\rm in}_{(j)}\frac{d\tau_{(j)}}{dt}u^{a}_{(j)}\delta_{d-1}(\mathbf{x}-\mathbf{y}_{(j)}(t))\Theta(-t)
\end{equation}
where $u^{a}_{(i,j)}$ are the $4$-velocities of the particles.
Conservation of $J^{a}$ implies conservation of charge at the interaction vertex
\begin{equation}
\sum_{j,in} q^{\rm in}_{(j)}=\sum_{i,out} q^{\rm out}_{(i)}.
\end{equation}
For $d \geq 5$, to leading order in $1/r$, the retarded solution to \eqref{elwaveeq} with the above current source is given by 
\begin{eqnarray}
A^{a} &=& 4\pi\int{d^{d}x'G^{a}{}_{a'}(x,x')J^{a'}(x')} \nonumber \\
&=& \frac{2\sqrt{\pi}(\alpha^{a}-\beta^{a})}{(2\pi r)^{\frac{d}{2}-1}}\bigg (\frac{\partial}{\partial U}\bigg)^{\frac{d-5}{2}} \bigg(\frac{\Theta(U)}{\sqrt{U}}\bigg)
\label{ASv}
\end{eqnarray}
where 
\begin{equation}
\begin{aligned}
\alpha^{a}(\hat{\mathbf{r}})=\sum_{(i)out}\frac{d\tau_{(i)}}{dt}\Big(\frac{q^{\rm out}_{(i)}u^{a}_{(i)}}{1-\hat{\mathbf{r}}\cdot\mathbf{v}_{(i)}}\Big)\\
\beta^{a}(\hat{\mathbf{r}})=\sum_{(j)in}\frac{d\tau_{(j)}}{dt}\Big(\frac{q^{\rm in}_{(j)}u^{a}_{(j)}}{1-\hat{\mathbf{r}}\cdot\mathbf{v}_{(j)}}\Big)\\
\end{aligned}
\end{equation}
and the parallel transport of the $u^{a}$'s is understood. The field tensor $F_{ab}=\nabla_{[a}A_{b]}$ is, to leading order, 
\begin{equation}
F_{ab}=\frac{2\sqrt{\pi}K_{[a}(\alpha_{b]}-\beta_{b]})}{(2\pi r)^{\frac{d}{2}-1}}\bigg (\frac{\partial}{\partial U}\bigg)^{\frac{d-3}{2}} \bigg(\frac{\Theta(U)}{\sqrt{U}}\bigg).
\label{retF}
\end{equation}
Again, for $d=3$, although the retarded solution for $A_a$ does not exist, there is a well-defined retarded solution for $F_{ab}$ and \eqref{retF} can be shown to also be valid for $d=3$.
We have
\begin{equation}
K^{[a}(\alpha^{b]}-\beta^{b]})=\sum_{(i)in}q^{\rm in}_{(i)} \frac{d\tau_{(i)}}{dt}\Big(\frac{K^{[a}q^{b]c}u_{(i)c}}{1-\hat{\mathbf{r}}\cdot\mathbf{v}_{(i)}}\Big) - \sum_{(j) out}q^{\rm out}_{(j)} \frac{d\tau_{(j)}}{dt}\Big(\frac{K^{[a}q^{b]c}u_{(j)c}}{1-\hat{\mathbf{r}}\cdot\mathbf{v}_{(j)}}\Big)
\end{equation}
where $q_{ab}$ is the projector onto directions tangent to spheres at large $r$.

The force acting on a test particle of charge $\mathcal{Q}$ with four-velocity $V^{a}$ is 
\begin{equation}
f^{a}=\mathcal{Q}F^{ab}V_{b}. \label{elforceeq}
\end{equation}
It follows immediately that the effects on test particle motion are very similar to the scalar case, except that, for a test particle at rest, the force now acts tangentially to spheres rather than in the null radial direction $K^a$. However, the $U$-dependence of the force is the same as in the scalar case. Again, for $d=3$ we obtain an infinite momentum memory effect. For all $d \geq 5$, we find that for a smoothed out source, there is no momentum memory effect. In five dimensions, there is again an infinite displacement memory effect. For $d > 5$, there is no displacement memory effect.

\section{The Gravitational Memory Effect}\label{sec:retgrav}

We wish to consider the gravitational memory effect arising from linearized perturbations $h_{ab}$ from a particle scattering event in an odd-dimensional Minkowski background with metric $\eta_{ab}$. Our detector will consist of two point particles that are far away from the interaction point and are initially at rest. To determine the gravitational memory effect we compute the relative displacement of the particles over time by integrating the geodesic deviation equation. For $d=3$, there is no gravitational radiation---i.e., spacetime is locally flat outside of sources---so there cannot be a gravitational memory effect. We therefore restrict consideration to $d \geq 5$ in our analysis below.

In the Lorentz gauge, we have $\nabla^{a}{\bar h}_{ab}=0$, where 
\begin{equation}
{\bar h}_{ab} \equiv h_{ab}-\frac{1}{2}\eta_{ab}h
\end{equation}
is the ``trace-reversed'' metric perturbation, with $h = \eta^{ab}h_{ab}$. The linearized Einstein equation with stress-energy $T_{ab}$ is given by 
\begin{equation}
\eta^{ab}\partial_{a}\partial_{b}{{\bar h}_{cd}}=-16\pi T_{cd}.\label{gravwaveeq}
\end{equation}
The retarded solution to (\ref{gravwaveeq}) is given by 
\begin{equation}
{\bar h}_{ab}=16\pi\int{d^{d}x'G_{ab}{}^{a'b'}(x,x')T_{a'b'}(x')}\label{gammaretint}
\end{equation}
with
\begin{equation}
G_{ab}{}^{a'b'}(x,x')=\eta_{a}{}^{a'}\eta_{b}{}^{b'}G(x,x')
\end{equation}
where $G(x,x')$ is the scalar Green's function to (\ref{waveeq}) and $\eta_{a}{}^{a'}$ is the parallel propagator. 

We, again, consider a scattering process of the type shown in Fig.\ref{scatter}.
The stress tensor of the outgoing particles, each with rest mass $m^{\rm out}_{(i)}$, has the form
\begin{equation}
T_{ab}^{(out)}=\sum_{i,out}m^{\rm out}_{(i)}\frac{d\tau_{(i)}}{dt}u_{(i)a}u_{(i)b}\delta_{d-1}(\mathbf{x}-\mathbf{y}_{(i)}(t))\Theta(t)
\end{equation}
with a similar expression holding for the stress-energy of the incoming particles.
Conservation of stress energy, $\nabla^{a}T_{ab}=0$, implies conservation of $4$-momentum at the interaction vertex
\begin{equation}
\sum_{i,out}m_{(i)}^{\rm out}u^{a}_{(i)}=\sum_{j,in}m_{(j)}^{\rm in}u^{a}_{(j)}.\label{conservmom}
\end{equation}
Evaluating \eqref{gammaretint} and converting from ${\bar h}_{ab}$ to $h_{ab}$, we obtain to leading order in $1/r$,
\begin{equation}
h_{ab}(x)=\frac{8\sqrt{\pi}(\alpha_{ab}-\beta_{ab})}{(2\pi r)^{\frac{d}{2}-1}}\bigg(\frac{\partial}{\partial U}\bigg)^{\frac{d-5}{2}}\bigg(\frac{\Theta(U)}{\sqrt{U}}\bigg)\label{gammaSv}
\end{equation}
where
\begin{equation}
\begin{aligned}
\alpha_{ab}(\hat{\mathbf{r}})=\sum_{(j)out}\frac{d\tau^{(j)}}{dt}\Big(\frac{m^{\rm out}_{(j)}}{1-\hat{\mathbf{r}}\cdot\mathbf{v}^{(j)}}\Big)\bigg(u_{a}^{(j)}u_{b}^{(j)}+\frac{1}{d-2}\eta_{ab}\bigg)\\
\beta_{ab}(\hat{\mathbf{r}})=\sum_{(i)in}\frac{d\tau^{(i)}}{dt}\Big(\frac{m^{\rm in}_{(i)}}{1-\hat{\mathbf{r}}\cdot\mathbf{v}^{(i)}}\Big)\bigg(u_{a}^{(i)}u_{b}^{(i)}+\frac{1}{d-2}\eta_{ab}\bigg)
\end{aligned}
\end{equation}
where, again, the parallel transport of the $u^{a}$'s is understood. 

The linearized Riemann tensor is 
\begin{equation}
R_{abcd}= \partial_{a}\partial_{[d}h_{b]c} - \partial_{c}\partial_{[d}h_{b]a}. 
\end{equation}
Using (\ref{gammaSv}), we obtain
\begin{equation}
R_{abcd}=\frac{8\sqrt{\pi}}{(2\pi r)^{\frac{d}{2}-1}}K_{[a}\Delta_{b][c}K_{d]}\bigg(\frac{\partial}{\partial U}\bigg)^{\frac{d-1}{2}}\bigg(\frac{\Theta(U)}{\sqrt{U}}\bigg) \label{Riem}
\end{equation}
where\footnote{The corresponding formula for $\Delta_{ab}$ for even-dimensional particle scattering given in \cite{Tolish1} erroneously omitted the factor of $q_{cd}u^{c}u^{d}$ in the terms proportional to $q_{ab}$.} 
\begin{eqnarray}
\Delta_{ab} &=& \sum_{(i)out} \frac{d\tau_{(i)}}{dt}\Big(\frac{m^{\rm out}_{(i)}}{1-\hat{\mathbf{r}}\cdot\mathbf{v}_{(i)}}\Big)\bigg(q_{ac}u^{c}_{(i)}q_{bd}u^{d}_{(i)}-\frac{q_{cd}u^{c}_{(i)}u^{d}_{(i)}}{d-2}q_{ab}\bigg) \nonumber \\
&& -\sum_{(j)in}\frac{d\tau_{(j)}}{dt}\Big(\frac{m^{\rm in}_{(j)}}{1-\hat{\mathbf{r}}\cdot\mathbf{v}_{(j)}}\Big)\bigg(q_{ac}u^{c}_{(j)}q_{bd}u^{d}_{(j)}-\frac{q_{cd}u^{c}_{(j)}u^{d}_{(j)}}{d-2}q_{ab}\bigg)
\label{memorytensor}
\end{eqnarray}
where, again, $q_{ab}$ is the projector onto directions tangent to spheres at large $r$.

The relative motion of test particles moving on geodesics is described by the geodesic deviation equation. If $V^a$ is the 4-velocity of one of the particles and $\xi^a$ denotes the deviation describing the relative displacement to the (infinitesimally) nearby particle, then
\begin{equation}
V^{e}\nabla_{e}(V^{f}\nabla_{f}\xi^{a})=R^{a}{}_{bcd}V^{b}V^{c}\xi^{d}.
\end{equation}
If the particles move nonrelativistically, then $V^{a} \approx (\partial/\partial t)^{a}$ and the deviation vector is (nearly) purely spatial. The geodesic deviation equation for the spatial components, $\xi^i$, of $\xi^a$ becomes
\begin{equation}
\frac{d^{2}\xi^{i}}{dt^{2}}=-{R^i}_{0j0}\xi^{j}.
\label{geodesicdevmemory}
\end{equation}
Using (\ref{Riem}) and integrating twice, we obtain
\begin{equation}
\Delta\xi^{i}=\frac{2\sqrt{\pi}}{(2\pi r)^{\frac{d}{2}-1}}\Delta_{j}{}^{i}\bigg(\frac{\partial}{\partial U}\bigg)^{\frac{d-5}{2}}\bigg(\frac{\Theta(U)}{\sqrt{U}}\bigg)\xi^{j}.
\end{equation}
Thus, the relative displacement, $\Delta \xi^i$, has exactly the same $U$-dependence as the momentum change, $\Delta P^a$, in the scalar and electromagnetic cases [see \eqref{scalarmom}]. Again, we must smooth out the source in the manner explained in the discussion of the scalar case in order to maintain consistency with our assumption of nonrelativistic 
motion, since without smoothing there will be arbitrarily large relative accelerations at small, positive $U$. With a smoothed source, there is no gravitational memory effect.

\section{Memory in the Slow Motion Limit}
To gain further insight into the absence of a gravitational memory effect in odd dimensions, we examine scalar, electromagnetic, and gravitational radiation in the slow motion limit of a source. We restrict consideration to $d \geq 5$. 

By \eqref{Greensfunctiontimederiv},
the retarded solution for a scalar field with a source $S(t', \mathbf{x}')$ of compact spatial support takes the form 
\begin{equation}
\phi(t,\mathbf{x}) = \frac{4\pi}{2\pi^{\frac{d-1}{2}}}\int{\bigg(-\frac{1}{2(t-t')}\frac{\partial}{\partial t'}\bigg)^{\frac{d-3}{2}}\bigg(\frac{\Theta(t-t'-|\mathbf{x}-\mathbf{x}'|)}{\sqrt{(t-t')^{2}-|\mathbf{x}-\mathbf{x}'|^{2}}}\bigg)}S(t', \mathbf{x}')d^{d}x'
\label{phigen}
\end{equation}
where, again, there is no need to do the ``$\epsilon$-regularization'' of \eqref{Greensfunctiontimederiv} since we are not interested in evaluating the solution within the support of $S$. 

We are interested in the leading order behavior of $\phi$ in $1/|\mathbf{x}|$ at fixed $U = t-|\mathbf{x}|$.
By definition of the derivative of a distribution, the $t'$-derivatives in \eqref{phigen} are to be evaluated by integration by parts. When the $t'$-derivative hits a factor of $1/(t-t') = 1/(U + |\mathbf{x}| -t')$, it produces a term of order $1/|\mathbf{x}|^2$, which does not contribute to the leading order behavior in $1/|\mathbf{x}|$.
Thus, to leading order in $1/|\mathbf{x}|$, we obtain
\begin{equation}
\phi(U,\mathbf{x}) = \frac{4\pi}{(2\pi)^{\frac{d-1}{2}}}\int{\frac{\Theta(U + |\mathbf{x}|-t'-|\mathbf{x}-\mathbf{x}'|)}{(U+ |\mathbf{x}|-t')^{\frac{d-3}{2}}\sqrt{(U + |\mathbf{x}| -t')^2-|\mathbf{x}-\mathbf{x}'|^2}}\partial_{t'}^{\frac{d-3}{2}}S(t', \mathbf{x}')d^{d}x'}.
\end{equation}
Now change variables from $t'$ to
\begin{equation}
T' = t' - |\mathbf{x}| + |\mathbf{x} - \mathbf{x}'|.
\end{equation}
We have
\begin{eqnarray}
\frac{1}{\sqrt{(U + |\mathbf{x}| -t')^2-|\mathbf{x}-\mathbf{x}'|^2}} &=&
\frac{1}{\sqrt{(U + |\mathbf{x} - \mathbf{x}' | -T')^2-|\mathbf{x}-\mathbf{x}'|^2}} \nonumber \\
&=& \frac{1}{\sqrt{U -T'+2 |\mathbf{x}-\mathbf{x}'|}}\frac{1}{\sqrt{(U -T')}} \nonumber \\
&\approx& \frac{1}{\sqrt{U -T'+2 |\mathbf{x}|}}\frac{1}{\sqrt{(U -T')}}.
\end{eqnarray}
Similarly, we have
\begin{equation}
\frac{1}{U+ |\mathbf{x}|-t'} = \frac{1}{U+ |\mathbf{x}- \mathbf{x}'|-T'} \approx \frac{1}{U+ |\mathbf{x}|-T'}.
\end{equation}
Hence, we obtain, to leading order in $1/|\mathbf x|$
\begin{eqnarray}
\phi(U,\mathbf{x}) = \frac{4\pi}{(2\pi)^{\frac{d-1}{2}}} && \int{\frac{1}{(U + |\mathbf{x}|-T')^{\frac{d-3}{2}}\sqrt{U-T'+2|\mathbf{x}|}}\bigg(\frac{\Theta(U-T')}{\sqrt{U-T'}}\bigg)} \times \nonumber \\
&& \times \,\, \partial_{T'}^{\frac{d-3}{2}}S(T' +  |\mathbf{x}| -  |\mathbf{x} - \mathbf{x}'|, \mathbf{x}')dT'd^{d-1} \mathbf{x}'.
\label{phiint1}
\end{eqnarray}
Since $|\mathbf{x}| \gg |\mathbf{x}'|$, we may write 
\begin{equation}
  |\mathbf{x}| -  |\mathbf{x} - \mathbf{x}'| \approx  \hat{\mathbf{x}}\cdot\mathbf{x}'
\end{equation}
where $\hat{\mathbf{x}} \equiv \mathbf{x}/|\mathbf{x}|$, so the argument of $S$ in the integral \eqref{phiint1} may be taken to be $S(T'+ \hat{\mathbf{x}}\cdot\mathbf{x}', \mathbf{x}')$.

A multipole expansion formula for $\phi$ can be obtained by formally expanding $S$ in a Taylor series in its first argument,
\begin{equation}
S(T'+ \hat{\mathbf{x}}\cdot\mathbf{x}', \mathbf{x}') = \sum_{n=0}^\infty \frac{1}{n!} ( \hat{\mathbf{x}}\cdot\mathbf{x}')^n \partial^n_{T'} S(T',\mathbf{x}').
\label{tay}
\end{equation}
Substituting this expansion into \eqref{phiint1} and formally interchanging summation and integration, we obtain
\begin{eqnarray}
\phi(U,\mathbf{x}) = \frac{4\pi}{(2\pi)^{\frac{d-1}{2}}} && \sum_{n=0}^\infty \frac{1}{n!} \int{dT'\frac{1}{(U + |\mathbf{x}|-T')^{\frac{d-3}{2}}\sqrt{U-T'+2|\mathbf{x}|}}\bigg(\frac{\Theta(U-T')}{\sqrt{U-T'}}\bigg)} \times \nonumber \\
&& \times \,\, \partial_{T'}^{n+\frac{d-3}{2}} \int ( \hat{\mathbf{x}}\cdot\mathbf{x}')^n S(T', \mathbf{x}')d^{d-1} \mathbf{x}'.
\label{phiint2}
\end{eqnarray}
We will not concern ourselves here with the precise conditions that $S$ must satisfy for the series formula \eqref{phiint2} to be valid, since our interest is in the ``slow motion limit,'' wherein $|\partial^{(n+1)}_{T'} S| \ll |\partial^n_{T'} S| /R$ throughout the source, where $R$ denotes the spatial extent of the source. In that case, it is clear that the dominant contribution will be obtained by keeping only
the lowest order ($n=0$) term on the right side of \eqref{tay} [or, equivalently, replacing $S(T'+ \hat{\mathbf{x}}\cdot\mathbf{x}', \mathbf{x}')$ by $S(T', \mathbf{x}')$ in \eqref{phiint1}]. We thereby obtain
\begin{equation}
\phi(U,\mathbf{x}) = \frac{4\pi}{(2\pi)^{\frac{d-1}{2}}}\int{dT'\frac{1}{(U + r -T')^{\frac{d-3}{2}}\sqrt{U-T'+2r}}\frac{\Theta(U-T')}{\sqrt{U-T'}}\partial_{T'}^{\frac{d-3}{2}}\Omega(T')}
\label{phiint3}
\end{equation}
where we have again written $r=|\mathbf{x}|$, and $\Omega(T')$ is the monopole moment of the source,
\begin{equation}
\Omega(T') = \int S(T', \mathbf{x}') d^{d-1} \mathbf{x}'.
\end{equation}
If we assume that, for some $T_0$, the time variation of $\Omega(T')$ is negligibly small for all $T' < T_0$, then for $r \gg U, |T_0|$, we have
\begin{equation}
\phi(U,\mathbf{x}) = \frac{2\sqrt{\pi}}{(2\pi r)^{\frac{d}{2}-1}}\int_{T_0}^U {dT'\frac{1}{\sqrt{U-T'}}\partial_{T'}^{\frac{d-3}{2}}\Omega(T')}.
\label{phiint4}
\end{equation}
If we further assume that, for some $T_1$, the time variation of $\Omega(T')$ is also negligibly small for $T'>T_{1}$, we see that $\phi(U,\mathbf{x})\rightarrow 0$ as $U\rightarrow \infty$.

For electromagnetic radiation, we obtain a similar expression for the components $A_\mu$ of the vector potential in Lorenz gauge to lowest order in $1/|\mathbf{x}|$ in the slow motion limit
\begin{equation}
A_{\mu}(U,\mathbf{x}) = \frac{4\pi}{(2\pi)^{\frac{d-1}{2}}}\int{dT'\frac{1}{(U + r -T')^{\frac{d-3}{2}}\sqrt{U-T'+2r}}\frac{\Theta(U-T')}{\sqrt{U-T'}}
\partial_{T'}^{\frac{d-3}{2}}\int{d^{d-1}\mathbf{x}'J_{\mu}(T',\mathbf{x}')}}.
\end{equation}
Using conservation of the charge current, $\partial_a J^{a}=0$, we find that the spatial components of the current satisfy
\begin{eqnarray}
\int{J^{i}d^{d-1}\mathbf{x}'} &=& \int{J^{j}\partial_{j}x'^{i} d^{d-1}\mathbf{x}'} \nonumber \\
&=&- \int{\partial_j J^{j}x'^{i}d^{d-1}\mathbf{x}'} \nonumber \\
&=&\partial_{T'}\int{J^{0}x'^{i}d^{d-1}\mathbf{x}'}=\partial_{T'}p^{i} 
\end{eqnarray}
where $p^{i}$ is the electric dipole moment of the source. Thus, we obtain, for the spatial components of the vector potential
\begin{equation}
A_{i}(U,\mathbf{x}) = \frac{4\pi}{(2\pi)^{\frac{d-1}{2}}}\int{dT'\frac{1}{(U + r-T')^{\frac{d-3}{2}}\sqrt{U-T'+2r}}\frac{\Theta(U-T')}{\sqrt{U-T'}}
\partial_{T'}^{\frac{d-1}{2}}p_i(T')}.
\end{equation}
The time component, $A_0$, of $A_\mu$ can then be obtained from the Lorenz gauge condition, $\partial_t A_0 = \partial_i A^i$.
If we assume that, for some $T_0$, the time variation of $\partial_{T'}p_i(T')$ is negligibly small for all $T' < T_0$---as will be the case if the source moves inertially at early times---then for $r \gg U, |T_0|$, we have
\begin{equation}
A_{i}(U,\mathbf{x}) = \frac{2\sqrt{\pi}}{(2\pi r)^{\frac{d}{2}-1}}\int_{T_0}^U {dT'\frac{1}{\sqrt{U-T'}}\partial_{T'}^{\frac{d-1}{2}}p_i(T')}.
\end{equation}
If we further assume that $\partial_{T'}p_{i}(T')$ is negligibly small for $T'>T_{1}$, we see that $A_{i}(U,\mathbf{x})\rightarrow 0$ as $U\rightarrow \infty$. 

Similarly, in the linearized gravitational case, we find that the spatial components of the ``trace-reversed'' metric perturbation, ${\bar h}_{ij}$, are given by
\begin{equation}
{\bar h}_{ij}(U, \mathbf{x})=\frac{16\pi}{(2\pi)^{\frac{d-1}{2}}}\int{dT'\frac{1}{(U + r-T')^{\frac{d-3}{2}}\sqrt{U-T'+2r}}\frac{\Theta(U-T')}{\sqrt{U-T'}}
\partial_{T'}^{\frac{d-3}{2}} \int{d^{d-1}\mathbf{x}'T_{ij}}}.
\end{equation}
Using conservation of stress energy $\partial_{a}T^{ab}=0$, we can rewrite the spatial integral as
\begin{equation}
\int{d^{d-1}\mathbf{x}'T^{ij}} = \frac{1}{2}\frac{\partial^{2}}{\partial T'^{2}}\int{d^{d-1}\mathbf{x}'T^{00}x'^{i}x'^{j}}=\frac{1}{2}\frac{\partial^{2}I_{ij}}{\partial T'^{2}}
\end{equation}
where $I_{ij}(T')$ is the moment of inertia tensor. 
Therefore we obtain
\begin{equation}
{\bar h}_{ij}(U, \mathbf{x})=\frac{8\pi}{(2\pi)^{\frac{d-1}{2}}} 
\int{dT'\frac{1}{(U + r-T')^{\frac{d-3}{2}}\sqrt{U-T'+2r}}\frac{\Theta(U-T')}{\sqrt{U-T'}}
\partial_{T'}^{\frac{d+1}{2}}I_{ij}(T')}.\label{lowestorderslowsource}
\end{equation}
The time components of ${\bar h}_{ab}$ can then be obtained from $\partial_t {\bar h}_{0 \mu} = \partial_i {{\bar h}^i}_\mu$.
If we assume that, for some $T_0$, the time variation of $\partial^3_{T'} I_{ij}(T')$ is negligibly small for all $T' < T_0$ (as will be the case if the source moves inertially at early times) then for $r \gg U, |T_0|$, we have
\begin{equation}
{\bar h}_{ij}(U,\mathbf{x}) = \frac{4\sqrt{\pi}}{(2\pi r)^{\frac{d}{2}-1}}\int_{T_0}^U {dT'\frac{1}{\sqrt{U-T'}}\partial_{T'}^{\frac{d+1}{2}}I_{ij}(T')}.
\label{barh}
\end{equation}

Now suppose that the source also satisfies $\partial^3_{T'} I_{ij}(T') \approx 0$ for all $T' > T_1$, as will be the case if the source moves inertially at late times. It then follows immediately from \eqref{barh} that ${\bar h}_{ij} (U) \to 0$ as $U \to \infty$. Since we also have ${\bar h}_{ij} (U) \to 0$ as $U \to -\infty$, we have $\Delta {\bar h}_{ij} = 0$. Carrying through the same type of analysis as given at the end of Sec. IV, we find that the memory effect is given by \cite{Tolish1}
\begin{equation}
\Delta \xi^{i} = \frac{1}{2}\mathcal{P}[\Delta h]_{j}{}^{i}\xi^{j}
\end{equation}
where
\begin{equation}
\mathcal{P} [\Delta h]_{ij} \equiv (q_{i}{}^{m}q_{j}{}^{n}-\frac{1}{d-2}q_{ij}q^{mn})(\Delta h)_{mn} =(q_{i}{}^{m}q_{j}{}^{n}-\frac{1}{d-2}q_{ij}q^{mn})(\Delta {\bar h})_{mn} 
\end{equation}
where $q_{ij}$ is the projector onto directions tangent to spheres at large $r$.
Thus, since $\Delta {\bar h}_{ij} = 0$, there is no memory effect in the slow motion limit for all odd $d \geq 5$.

\section{Summary}

Putting together the results of this paper for odd $d$ with the results of \cite{Tolish1} for even $d$, we may summarize the results for the memory effect at order $1/r^{d/2-1}$ resulting from particle scattering as follows:

\begin{itemize}

\item For spacetime dimension $d=3$, in the scalar and electromagnetic cases there is an infinite momentum memory effect on test particles. There is no gravitational radiation for $d=3$, so there is no gravitational memory effect.

\item For spacetime dimension $d=4$, there is a nontrivial scalar and electromagnetic momentum memory
effect, i.e., a test particle will receive a finite momentum kick from the emitted radiation. There is a nontrivial gravitational memory effect, i.e., the relative configuration of test particles will be permanently displaced by a finite amount by the emitted radiation.

\item For spacetime dimension $d=5$, there is no scalar or electromagnetic momentum memory effect, but a test particle will be displaced by an infinite amount by the emitted radiation. There is no gravitational memory effect, i.e., the relative configuration of test particles will return to its original relative configuration after passage of the emitted radiation.

\item For spacetime dimension $d=6$, there is no scalar or electromagnetic momentum memory effect, but, for a "smoothed out" source, a test particle will be displaced by a finite amount by the emitted radiation. The amount of the displacement will depend on the details of the smoothing. There is no gravitational memory effect.

\item For spacetime dimension $d>6$, there is no scalar, electromagnetic, or gravitational memory effect of any kind.

\end{itemize}

\bigskip

\noindent
{\bf Acknowledgements}

We wish to thank Alex Tolish for helpful discussions. This research was supported in part by NSF Grant No. PHY 15-05124 to the University of Chicago.

\end{document}